\documentclass[journal]{IEEEtran}
\usepackage{soul}
\usepackage{cite}
\usepackage{amsmath,amsfonts,amsxtra,amssymb,latexsym,amscd,amsthm,mathrsfs,bm}
\usepackage{algorithmic}
\usepackage{algorithm}
\usepackage{graphics}
\usepackage{graphicx}
\usepackage{hyperref}
\usepackage{textcomp}
\usepackage{balance}
\usepackage{booktabs}
\usepackage{caption}
\usepackage{subcaption}
\usepackage{cuted, nccmath}
\usepackage{amsmath}
\usepackage{lipsum}
\usepackage[dvipsnames]{xcolor}
\usepackage[utf8]{inputenc}
\usepackage[english]{babel}
\usepackage{filecontents}
\usepackage{xcolor}
\usepackage{thmtools}
\usepackage{multirow}

\usepackage{stackengine}
\usepackage{stfloats}

\usepackage{caption}
\usepackage[compact]{titlesec}  
\titlespacing{\section}{0pt}{3pt}{3pt}
\titlespacing{\subsection}{0pt}{2pt}{2pt}
\makeatletter
\renewcommand{\fnum@figure}{Fig. \thefigure}
\makeatother

\DeclareMathOperator{\E}{\mathbb{E}}

\newcommand{\EX}[1]{\E\left\{{#1}\right\}}

\newcommand{\CG}[2]{\mathcal{CN}\left({#1},{#2}\right)}

\newcommand{\dl}{\mathrm{d}}

\newtheoremstyle{mythm}%
{3pt}
{3pt}
{}
{}
{\bfseries}
{}
{.5em}
{}%

\declaretheoremstyle[
headpunct=\textup{:},
bodyfont=\normalfont,
]{myremark}
\theoremstyle{myremark}
\newtheorem{remarkex}{Remark}

\newenvironment{remark}
{\pushQED{\qed}\remarkex}
{\popQED\endremarkex}

\ifCLASSINFOpdf
\else
\fi
\begin{document}
%
\title{\huge Distributed Graph Neural Network Design for Sum Ergodic Spectral Efficiency Maximization in Cell-Free Massive MIMO }
%
%

\author{Nguyen~Xuan~Tung, Trinh~Van~Chien, \textit{Member}, \textit{IEEE}, Hien Quoc Ngo, \textit{Senior Member}, \textit{IEEE}, Won~Joo~Hwang,~\textit{Senior Member},~\textit{IEEE} 
    \thanks{Nguyen Xuan Tung is with the Department of Information Convergence Engineering, Pusan National University, Busan 46241, Republic of Korea (email: {tung.nguyenxuan1310@pusan.ac.kr}).}
    \thanks{Trinh Van Chien is with the School of Information and Communication Technology (SoICT), Hanoi University of Science and Technology (HUST), Vietnam (email: chientv@soict.hust.edu.vn).}
     \thanks{Hien Quoc Ngo is with the Centre for Wireless Innovation (CWI), Queen's University Belfast, U.K. (hien.ngo@qub.ac.uk).}  
    \thanks{Won-Joo Hwang is with the School of Computer Science and Engineering, Center for Artificial Intelligence Research, Pusan National University, Busan 46241, South Korea (e-mail: wjhwang@pusan.ac.kr).}
}

%
%

\markboth{}
{Shell \MakeLowercase{\textit{et al.}}: Distributed graph neural network based framework for cell-free massive-MIMO}
%



\maketitle

\makeatletter
\def\ps@IEEEtitlepagestyle{
  \def\@oddfoot{\mycopyrightnotice}
  \def\@evenfoot{}
}
\def\mycopyrightnotice{
  {\footnotesize
  \begin{minipage}{\textwidth}
  \centering
  Copyright~\copyright~2024 IEEE. Personal use of this material is permitted. However, permission to use \\this material for any other purposes must be obtained from the IEEE by sending a request to pubs-permissions@ieee.org.
  \end{minipage}
  }
}

\begin{abstract}
    This paper proposes a distributed learning-based framework to tackle the sum ergodic rate maximization problem in cell-free massive multiple-input multiple-output (MIMO) systems by utilizing the graph neural network (GNN). Different from centralized schemes, which gather all the channel state information (CSI) at the central processing unit (CPU) for calculating the resource allocation, the local resource of access points (APs) is exploited in the proposed distributed GNN-based framework to allocate transmit powers. Specifically, APs can use a unique GNN model to allocate their power based on the local CSI. The GNN model is trained at the CPU using the local CSI of one AP, with partially exchanged information from other APs to calculate the loss function to reflect system characteristics, capturing comprehensive network information while avoiding computation burden. Numerical results show that the proposed distributed learning-based approach achieves a sum ergodic rate close to that of centralized learning while outperforming the model-based optimization.
\end{abstract} 
\IEEEpeerreviewmaketitle
\section{Introduction}
Cell-free massive multiple-input multiple-output (MIMO) characterizes a network configuration wherein numerous access points (APs) coherently serve multiple user equipments (UEs) with the same time-frequency resources. This paradigm holds the potential to yield network enhancements in throughput, reliability, and energy efficiency \cite{CellFree_vs_SmallCell_hien2017, EECellFree_hien2018}. To fully exploit the benefits of cell-free massive MIMO systems, numerous studies and investigations have been conducted in various subjects \cite{Utility_Maximization_Farooq2021, ScalableCellFree_Bjornson2020}. Cell-free massive MIMO systems pose increasing challenges as network size grows, making conventional optimization-driven methodologies unsuitable due to time constraints and practical implementation feasibility.

Deep learning (DL) is a transformative technology for cell-free massive MIMO systems \cite{DLbasedPowerControl_Rajapaksha2021, LearningPowerControl_Zaher2023}. However, DL models are context-specific, operating effectively only within the scenarios present during training. Consequently, retraining is needed when network dimensions change, limiting flexibility. Graph Neural Networks (GNNs) offer a solution to this scalability challenge by modeling wireless systems as graphs, where nodes and edges represent system entities and their interconnections. In this way, a GNN forecasts the outcomes on nodes or edges by analyzing their inherent features. Therefore, GNNs depend on the attributes of nodes and edges rather than system dimensions.
Despite the popularity of GNNs in wireless systems \cite{OverviewGNNforWireNet_Shiwen2021, SurveyGNNmeetWireComm_Lee2022, Giang10738316}, there have been few studies on their application in cell-free massive MIMO. In \cite{CentraliedGNN_Yifei2022}, a heterogeneous GNN-based message passing was proposed for power control in an uplink cell-free massive MIMO system, considering UEs and APs as different types of entities in the represented graph. The paper  \cite{PowerControlGNN_Globe_Salaun2022} developed a GNN to optimize the downlink max-min power allocation, incorporating maximum ratio transmission beamforming. Both \cite{CentraliedGNN_Yifei2022} and \cite{PowerControlGNN_Globe_Salaun2022} use a centralized GNN model at the central processing unit (CPU), assuming full CSI of the entire system is available at the CPU. While these approaches achieve high performance, they incur substantial computational and communication overhead, and the computational capacity of individual APs is underutilized. Previous distributed deep learning approaches, like \cite{Distributed_cfmMIMO_DNN_Zaher2022}, reduce the central server's burden but are limited to specific training configurations, restricting generalizability. Meanwhile, distributed optimization approaches, like \cite{Li_You_Distributed_Optimization}, still require information exchange between APs, increasing latency and complexity, especially in large-scale networks.


To resolve the above issue, we propose a novel distributed GNN-based framework where each AP leverages the local CSI to predict power allocation coefficients. During the training phase, the GNN model is trained at the CPU using the local CSI of a single AP, while combined information from other APs derived from their CSI is required to calculate the loss function. This combined information reflects desired signal, pilot contamination, and user interference, allowing the model to capture comprehensive network characteristics while reducing the computational burden. To the best of our knowledge, this is the first research deploying GNN in a distributed manner to solve a power allocation problem in cell-free massive MIMO. Our main contributions are briefly summarized as follows: \emph{i}) We study a  distributed GNN to allocate the transmit powers in the downlink cell-free massive MIMO. Each AP uses the proposed GNN model to predict its power coefficient based on its local channel statistics; \emph{ii}) During the training phase, we propose a procedure to update the GNN model and exchange information between the CPU and APs to enhance the learning ability and share the computational burden; and \emph{iii}) Numerical results show our distributed GNN framework performs nearly as well as centralized learning and outperforms optimization-based methods.


\emph{Notation:} Boldface lowercase letters and boldface uppercase letters denote vectors and matrices, respectively. Let $()^T$ and $()^*$ denote the transpose and conjugate, respectively. The Euclidean norm and expectation are denoted by $\|\cdot\|$, and $\mathbb{E}\{\cdot\}$. Finally, we define the circularly symmetric complex Gaussian distribution with variance $\sigma^2$ by $\mathcal{CN}(0,\sigma^2)$.

\section{Downlink Cell-free Massive MIMO} \label{SystemModel}
\subsection{System and Signal Models}
We consider a cell-free massive MIMO system where a set of UEs, $\mathcal{N} = \{1,..., N\}$, are jointly served by a set of APs, $\mathcal{K} = \{1,...,K\}$, simultaneously. Each AP has $M$ antennas, while each UE is equipped with a single antenna. All the $K$ APs are connected to a CPU \cite{CellFree_vs_SmallCell_hien2017} for resource management and signal processing. APs estimate the propagation channels in the uplink pilot training phase. The minimum mean-square error estimate (MMSE) of the channel between the $k$-th AP and the $n$-th UE is given by

\begin{equation}
    \begin{aligned}
        \hat{\mathbf{h}}_{kn} &= \mathbb{E}\{ \mathbf{h}_{kn}\tilde{\mathbf{y}}_{p,kn}^H\}
        (\mathbb{E}\{\tilde{\mathbf{y}}_{p,kn}\tilde{\mathbf{y}}_{p,kn}^H\})^{-1} \tilde{\mathbf{y}}_{p,kn} \\
        &=\frac{\sqrt{\tau_p \rho_p} \varsigma_{kn}}{\tau_p \rho_p \sum\limits_{n'\in \mathcal{N}} \varsigma_{kn'} |\boldsymbol{\theta}_n^H \boldsymbol{\theta}_{n'}|^2 + 1}\tilde{\mathbf{y}}_{p,kn},
    \end{aligned}
\end{equation}
where $\tilde{\mathbf{y}}_{p,kn} = \boldsymbol{Y}_{p,k}\boldsymbol{\theta}_n$, with $\boldsymbol{Y}_{p,k} = \sqrt{\tau_p \rho_p} \sum\limits_{n\in \mathcal{N}} \mathbf{h}_{kn} \boldsymbol{\theta}_n^H + \boldsymbol{\Xi}_{p,k}$ is the received pilot signal at the $k$-th AP. Here, $\mathbf{h}_{kn} \in \mathbb{C}^{M\times 1}$ is the channel between the $k$-th AP and the $n$-th UE, and its elements are independent and identically distributed (i.i.d.) $\mathcal{CN}(0,\varsigma_{kn})$ with $\varsigma_{kn}$ denoting the large-scale fading coefficient. Meanwhile, $\tau_p$, $\rho_p$, $\boldsymbol{\theta}_n$, and $\boldsymbol{\Xi}_{p,k}$ are the uplink training duration, the normalized signal-to-noise ratio (SNR) of each pilot symbol, the pilot sequence used by the $n$-th UE, and the additive noise at the $k$-th AP, respectively.

Given the channel estimate obtained from the pilot training phase, the $k$-th AP transmits the precoded signal, $\mathbf{x}_k = \sqrt{\rho_d} \sum_{n\in\mathcal{N}} \sqrt{P_{kn}} \hat{\mathbf{h}}^*_{kn} s_n $, to the UEs.  Here, $\rho_d$, $s_n$ with $\EX{|s_n|^2}=1$ and $P_{kn}$ are the maximum normalized downlink SNR, the symbol intended for the $n$-th UE and the corresponding allocated power, respectively. Due to the limited power budget, the total power at the $k$-th AP is constrained as

\begin{align}
\label{eq:pct}
\sum\limits_{n\in \mathcal{N}} P_{kn}v_{kn} \leq 1/M, \quad \forall k\in \mathcal{K},
\end{align}
where $v_{kn}\triangleq \mathbb{E}\{|[\hat{\mathbf{h}}_{kn}]_m |^2\} = \frac{{\tau_p \rho_p} \varsigma^2_{kn}}{\tau_p \rho_p \sum_{n'\in \mathcal{N}} \varsigma_{kn'} |\boldsymbol{\theta}_n^H \boldsymbol{\theta}_{n'}|^2 + 1}$. The signal received at the $n$-th UE is given by
\begin{align}\label{eq:rk1}
y_{\dl,n} 
&= 
\sum_{k\in \mathcal{K}}\mathbf{h}^T_{kn}\mathbf{x}_{k} +  
\xi_{\dl,n}\nonumber\\
 &=
\sqrt{\rho_d}\sum_{k\in\mathcal{K}}\sum_{n'\in \mathcal{N}}
\sqrt{P_{kn'}}\mathbf{h}^T_{kn}\hat{\mathbf{h}}_{kn'}^* s_{n'} + \xi_{\dl,n},
\end{align}
where $\xi_{\dl,n} \sim \CG{0}{1}$ is additive noise at the $n$-th UE. Note that the desired signal $s_n$ is detected from $y_{\dl,n}$ in \eqref{eq:rk1}. The closed-form expression of the downlink ergodic rate of the $n$-th user is adopted from \cite{EECellFree_hien2018} as expressed in \eqref{eq:Theo_rateexpr1}.
\begin{figure*}[b]
\hrulefill
\begin{align}
\label{eq:Theo_rateexpr1}
R^{\text{dl}}_{n}
 =
 \log_2
    \left(
    1 + \frac{\rho_d M^2\left(\sum\limits_{k\in \mathcal{K}} \sqrt{P_{kn}}v_{kn} \right)^2 }{ \rho_d M^2\sum\limits_{n'\neq n, n' \in \mathcal{N}}\left(\sum\limits_{k\in \mathcal{K}} 
    \sqrt{P_{kn'}}v_{kn'}{\varsigma_{kn}}/{\varsigma_{kn'}} \right)^2| \pmb{\theta}_{n'}^H\pmb{\theta}_{n}|^2 + \rho_dM\sum\limits_{n'\in \mathcal{N}}\sum\limits_{k\in \mathcal{K}} P_{kn'}v_{kn'}\varsigma_{kn} +1 }
    \right),
\end{align}
\end{figure*}
\subsection{Problem Formulation}
We aim to optimally allocate the power control coefficients $P_{kn}$ to maximize the total downlink ergodic rate of all the UEs subject to the limited power budget at the APs. The optimization problem is mathematically formulated as follows
\begin{subequations}\label{eq opt 1}
    \begin{align}
        && \underset{\substack{ \{P_{kn}\} }}{\textrm{maximize}} & \quad \sum\limits_{n\in \mathcal{N}} R^\text{dl}_{n} \\
        &&\textrm{subject to}&\quad\sum_{n\in\mathcal{N}} P_{kn}v_{kn} \leq 1/M, \forall  k\in\mathcal{K} \label{eq opt 1 - constraint1}\\
        &&& \quad P_{kn} \geq 0, \forall n\in \mathcal{N}, \forall k\in \mathcal{K}. \label{eq opt 1 - constraint2} 
    \end{align}
\end{subequations}
We emphasize that the problem in \eqref{eq opt 1} is nonconvex due to its objective function with the ergodic rate in \eqref{eq:Theo_rateexpr1}. This problem can be solved by using sequential convex optimization. However, this scheme is performed in a centralized manner with high computational complexity \cite{EECellFree_hien2018}. Consequently, this motivates us to seek a better-performance solution with reduced complexity, using prior information on the system topology.


\section{Distributed GNN-Based Learning Algorithm}
In this section, we construct a novel GNN-based learning algorithm to solve the optimization problem in \eqref{eq opt 1} in a distributed fashion.

\subsection{Cell-Free Massive MIMO as Multiple Graphs} A graph's formal definition is given by a pair $G = \langle \mathcal{V},\mathcal{E} \rangle$, where $\mathcal{V}$ and $\mathcal{E}$ are the sets of nodes and edges, respectively. An edge $(i, j) \in \mathcal{E}$ depicts the relationship between nodes $i$ and $j$. The attribute of the $i$-th node is represented by $v_i \in \mathbb{R}^{\mathsf{dn}}$, while that of the edge $(i, j)$ is  $e_{ij}\in \mathbb{R}^{\mathsf{de}}$, where $\mathsf{de}$ and $\mathsf{dn}$ are the sizes of edge's and node's features, respectively. The adjacent set $\mathcal{N}_i = \{j|(j, i)\in \mathcal{E}\}$ denotes the set of neighbors of the $i$-th node.
\begin{figure}
    \includegraphics[trim=0cm 0cm 0cm 0cm, clip=true, width=3.3in]{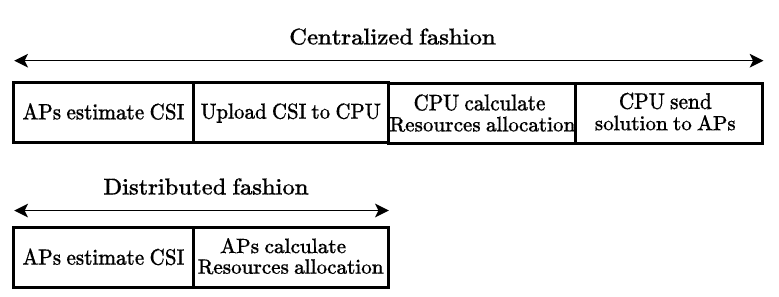}
    \caption{The centralized versus distributed fashion.}
    \label{Decen_vs_Centralized_Flowchart}
\end{figure}
\begin{figure}[t]
    \includegraphics[trim=0cm 0cm 0cm 0cm, clip=true, width=3.3in]{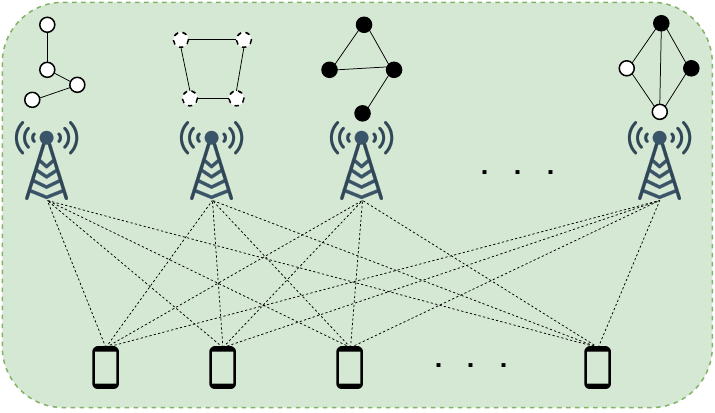}
    \caption{The cell-free massive MIMO as multiple graphs.}
    \label{CellFree_as_MultiGraphs}
\end{figure}
In a centralized vision, \cite{CentraliedGNN_Yifei2022}, a cell-free massive MIMO system was represented by a single graph. Therein, the CSI and power budgets of all the APs are gathered with the attributes of edges and nodes. A GNN model was proposed to learn the represented graph and predict resource allocation universally. This learning framework allows the trained GNN model to seek for the best solution in the testing phase. Nevertheless, it encounters a critical issue that the entire CSI should be collected at the CPU for training purposes with the increasing computational resources and fronthaul signaling.

To resolve the issue, we construct a distributed leaning-based framework to train and operate GNNs model locally. In more detail, every AP in the system uses a GNN model to predict the power coefficients based on its local CSI. The block diagram of the distributed power allocation in comparison to the centralization is illustrated in Fig.~\ref{Decen_vs_Centralized_Flowchart}.  The distributed scheme facilitates quick adaptation to system changes and leverages the computational capacities of individual APs.\footnote{The centralized scheme exhibits a slow response due to  exchanged information between the APs and the CPU to find a global optimum.} An AP only possesses the local information, thus the cell-free massive MIMO system is viewed as $|\mathcal{K}|$ isolated graphs as in Fig. \ref{CellFree_as_MultiGraphs}, where each AP forms a single graph. The constructed graph at the $k$-th AP contains only its connection to  the $|\mathcal{N}|$ UEs. We consider every link between an AP and a UE as a node. Thereby, the $k$-th graph is denoted by $G_k = \{\mathcal{V}_k, \mathcal{E}_k\}$, where $\mathcal{V}_k$ is the set of $|\mathcal{N}|$ signal nodes corresponding to $|\mathcal{N}|$ AP-UE links. Meanwhile, $\mathcal{E}_k$ is the set of edges representing the interference between signal nodes. The features of nodes and edges are defined as follows
\begin{itemize}
    \item \emph{Signal nodes}: There are $|\mathcal{V}_k| = |\mathcal{N}|$ signal nodes. The $n$-th node's feature of the $k$-th graph is $[\boldsymbol{Z}_{k}]_n = [\varsigma_{kn}, \boldsymbol{\theta}_{n}^T, \rho_d]$, and $\boldsymbol{Z}_k\in \mathbb{R}^{|\mathcal{N}| \times (|\tau_p| + 2)}$ is the feature matrix of the signal nodes.
    \item \emph{Interference edges}: There are $|\mathcal{E}_k| = |\mathcal{N}|\times |\mathcal{N}|$ interference edges. The edge feature between the two nodes $n$ and $n'$ of the $k$-th graph is $[\boldsymbol{E}_k]_{nn'} = [\varsigma_{kn}, \varsigma_{kn'}]$, otherwise $  [\boldsymbol{E}_k]_{nn'} = 0,\text{ if } n=n'$, and $\boldsymbol{E}_k \in \mathbb{R}^{|\mathcal{N}|\times |\mathcal{N}| \times 2}$ is the adjacency feature matrix.
\end{itemize}
\begin{remark}\label{Remark: Corperated training}
     To predict the power control coefficients in the testing phase, the GNN model at each AP requires only local channel statistics. Therefore, in this phase, the proposed framework allows the system to operate at APs without the need for CPU management. Due to the distributed nature of the framework, the interference links from other APs cannot be directly represented in the graph at each AP. Therefore, during the training phase, the CSI from other APs is requested so that the GNN model can indirectly learn their influence, capturing entire network behaviors and improving the learning policy. This approach ensures that the GNN model understands the effects of all APs, including desired signal, pilot contamination, and user interference, while using only local CSI during the testing phase.
\end{remark}


\subsection{GNN Designed based on the Message Passing Method}
To capture information from all the edges and nodes in the graph, the message-passing GNN (MPGNN) has been proposed. Each node receives the aggregated messages from its neighbors that contain the combined information from the node's neighbors and the corresponding edge's features. The forward computation of the $n$-th node in the $l$-th layer designed for the $k$-th AP is
\begin{equation}
    \begin{aligned}
        &{y}_{kn}^{(l)} = \phi_1 \Big(\text{CONCAT}([\boldsymbol{Z}_{k}]_n^{(l-1)}, \\
        & \quad \quad \quad\quad\quad \underset{n' \in \mathcal{N}_{(kn)}}{\text{AGG}} \{\phi_2 ([\boldsymbol{Z}_{k}]_{n'}^{(l-1)},[\boldsymbol{E}_k]_{n'n})\} )\Big),\\
        & {x}_{kn}^{(l)} = \text{ReLU}({y}_{kn}^{(l)}),
    \end{aligned}
\end{equation}
where $\phi_1$ and $\phi_2$ are two multi-layer perceptrons (MLPs) using the same at all MPGNN layers. Therein, $\phi_2$ yields the message that each node should pass to its neighbors. The neighbors' messages are aggregated via the aggregation function (AGG). A concatenation function (CONCAT) is utilized to concatenate the node's feature and the aggregated message from neighbors before passing through $\phi_1$ to infer the output. At the last layer, say the $L$-th layer, we set the activation function as
\begin{equation} \label{Activation fn: Power constraint}
P_{kn}^{L} = \frac{x_{kn}^{L}}{\sum_{n'=1}^N x_{kn'}^{L}v_{kn'}},
\end{equation}
to ensure the power coefficients satisfying the constraint \eqref{eq opt 1 - constraint1}. Moreover, the loss function is defined as
\begin{equation}\label{Loss fn: Sumrate}
        \mathcal{L}(\boldsymbol{\Psi}) = -\sum\nolimits_{n\in \mathcal{N}} \EX{R^\mathrm{dl}_{n}},
\end{equation}
where the downlink ergodic rate $R^\text{dl}_{n}$ is given in \eqref{eq:Theo_rateexpr1}, and $\boldsymbol{\Psi}$ is the GNN model parameters. Note that each AP only has its own data, therefore, the information sharing is needed to calculate the loss function in \eqref{Loss fn: Sumrate}.  We will hereafter propose a framework for sharing information between APs and the CPU during the training phase.
\subsection{Distributed GNN-based Learning}
\begin{figure}[t]
     \centering
    \includegraphics[trim=0cm 0cm 0cm 0cm, clip=true, width=3.5in]{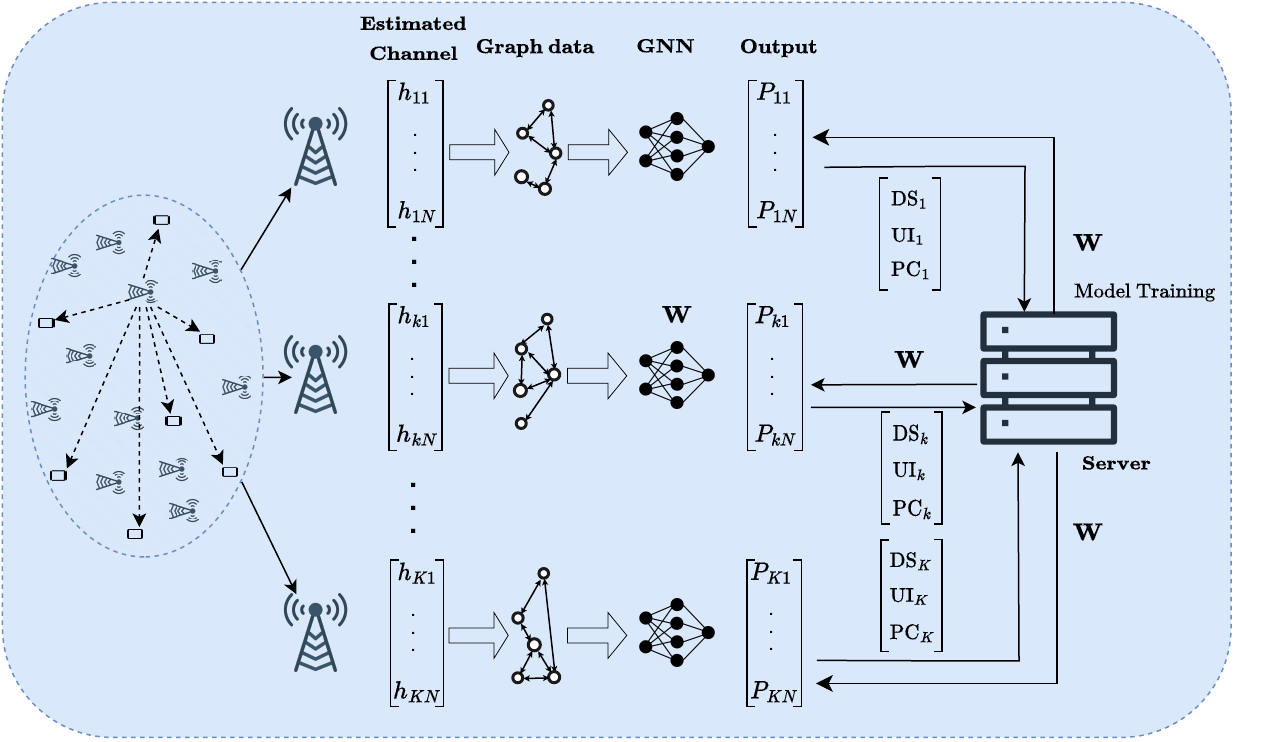}
    \caption{The general flow chart of the training phase.}
    \label{Fig: DecentralizedStructure}
\end{figure}

The APs are assumed to be independent and identically distributed. Therefore, we can train the GNN model based only on the local information of one AP. However, during the training phase, as discussed in Remark \ref{Remark: Corperated training}, the CPU will receive the channel statistics and the power coefficients from all APs to execute the entire computations to train a unique GNN model. This straightforward approach has a high cost due to the huge computation during the training phase. To robust the system, we suggest sharing the computation burden among the devices by requesting all APs to send the processed channel information instead of the raw data. Specifically, the $k$-th AP sends the following information:  $\text{DS}_k \in \mathbb{R}^{|\mathcal{N}|\times 1}$, where $[\text{DS}_k]_n = \sqrt{\rho_dP_{kn}}v_{kn}$; and  $\text{PC}_k \in \mathbb{R}^{|\mathcal{N}|\times |\mathcal{N}|}$, where $[\text{PC}_k]_{n'n} = |\sqrt{\rho_dP_{kn'}}v_{kn'}\frac{\varsigma_{kn}}{\varsigma_{kn'}} \pmb{\theta}_{n'}^H\pmb{\theta}_{n}|$; and $\text{UI}_k \in \mathbb{R}^{|\mathcal{N}|\times |\mathcal{N}|}$, where $[\text{UI}_k]_{nn'} = \rho_dP_{kn'}v_{kn'}\varsigma_{kn}$. 
The ergodic rate of the $n$-th UE can be computed as in \eqref{eq:1}.
\begin{figure*} 
\begin{equation} \label{eq:1}
    \begin{aligned}
    R_{n}^\text{dl}=\log_2
    \left(
    1 + \frac{M^2(\sum\limits_{k\in \mathcal{K}} [\text{DS}_k]_{n} )^2 }{ M^2\sum\limits_{ ^{n'\neq n,}_{n' \in \mathcal{N}}}(\sum\limits_{k\in \mathcal{K}} [\text{PC}_{k}]_{n'n})^2 + M\sum\limits_{n'\in \mathcal{N}}\sum\limits_{k\in \mathcal{K}} [\text{UI}_k]_{nn'} +1 }
    \right).
    \end{aligned}
\end{equation}
\hrule
\end{figure*} 
The GNN model at the CPU is trained given the synthesized information encompassing the desired signal, the pilot contamination, and the mutual interference. Such information is evaluated based on the CSI and power control coefficients among the APs. Note that the power coefficients at APs significantly influence the training performance. In the testing phase, all APs use the GNN model. Hence, the  GNN model will be trained under the influence of the previously shared information. Therefore, we suggest that the CPU will broadcast the trained GNN model to all APs frequently during the training phase. As a result, the GNN model can quickly evaluate its influence on the entire network, and then make necessary adjustments to optimize the performance. 
The detailed procedure is summarized in Algorithm \ref{Alg: DecentralizedGNN}, while the general framework is depicted in Fig. \ref{Fig: DecentralizedStructure}.
\begin{algorithm}[t]
    \caption{Distributed GNN for power allocation.}\label{Alg: DecentralizedGNN} 
    \begin{algorithmic}[1] 
      \STATE \textbf{Initialize}: The GNN model $\boldsymbol{\Psi}^{\text{curr}}$ and share to all APs.
        \REPEAT 
        \STATE All APs use $\boldsymbol{\Psi}^{\text{curr}}$ to predict power coefficients and transmit the processed channel information to the CPU.
        \STATE The CPU receives the processed channel information and use it to calculate loss. The GNN model is then updated, $\boldsymbol{\Psi}^{\text{upd}}$.
        \STATE The CPU broadcasts the latest model to all APs and set $\boldsymbol{\Psi}^{\text{curr}} = \boldsymbol{\Psi}^{\text{upd}}$.
        \UNTIL {Convergence.} 
        \STATE \textbf{Output}: The trained GNN at all APs.
    \end{algorithmic}
\end{algorithm} 
\begin{remark}
By requesting all APs to send the partial channel information consisting of $\text{DS}$, $\text{PC}$, and $\text{UI}$ to compute the loss function, the CPU requires the computational complexity in the order of $\mathcal{O}(KN^2)$, according to $[12]$. 
\end{remark}
\subsection{Data Exchange Analysis}
We assume that the size of the distributed GNN-based framework is $|\boldsymbol{\Psi}|$, and therefore, the exchanging parameters of both methods are summarized in Table~\ref{Table: Data Exchange Comparison}.
\begin{table}[t]
\centering
\caption{Information exchange comparison.}
\begin{tabular}{|c|c|c|c|c|} 
\hline
& \multicolumn{2}{|c|}{Distributed scheme} & \multicolumn{2}{|c|}{Centralized scheme} \\
\hline
& Uplink & Downlink & Uplink & Downlink \\
\hline
Training phase  & $K(N^2+N)$ & $K|\boldsymbol{\Psi}| $ & $K N$ & $K N$ \\ 
\hline 
Operating phase  & 0 & 0 & $K N$ & $K N$ \\ 
\hline
Training participants & \multicolumn{2}{|c|}{CPU and all APs} & \multicolumn{2}{|c|}{CPU} \\
\hline 
\end{tabular}\label{Table: Data Exchange Comparison}
\end{table}
The outstanding feature of the proposed distributed GNN can be seen as it operates without the need for data exchange between the APs and the CPU. In contrast, the centralized GNN requests all the APs to upload their local channel state information frequently to the CPU and wait for the optimized power coefficients. In the training phase, the distributed framework exchanges the information consisting of $|\boldsymbol{\Psi}|$ GNN model parameters and $K(N^2+N)$ parameters from the processed channel state information. Meanwhile, the centralized GNN must send the channel statistics from the APs to the CPU, and the power coefficients are sent back from the CPU to the APs. However, it is worth noting that training the GNN model at the CPU  results in a significant reduction in the workload at the APs. In addition, the centralized GNN encodes the whole network as a single graph, which becomes increasingly complex for large-scale systems. In contrast, the distributed GNN treats the system as multiple smaller graphs, enabling parallel processing across all the APs. As a result, the proposed distributed GNN provides superior computational speed as opposed to the centralized scheme, especially when dealing with a substantial number of APs and UEs. The running time will be shown in Section \ref{ExperimentalEvaluation} to illustrate the enhanced computational efficiency of our proposed approach.

\section{Performance Evaluation}\label{ExperimentalEvaluation}
This section evaluates the performance of the proposed distributed GNN-based framework by numerical results. For comparison, we further consider the following benchmarks: 1) \textbf{Centralized GNN-based framework} gathers full CSIs at the CPU to construct a represented graph for predicting the transmit power coefficients; 2) \textbf{Optimization-based approach} optimizes the transmit power coefficients by approximating the problem to a convex form using a slack variable and the first-order Taylor approximation, as described in [$2$, Section IV]; 3) \textbf{Proportional allocation} allocates power based on large-scale fading coefficients, with users having higher coefficients receiving more power; 4) \textbf{Equal allocation} distributes power equally among all users, regardless of their large-scale fading coefficients.


For parameter setting, we consider cell-free massive MIMO systems with the number of UEs $N=[5, 6, 10, 15, 20]$ and the number of APs $K = [20, 30, 40, 50]$. The locations of APs and UEs are uniformly distributed in an area of $1\text{km}\times1\text{km}$ squares. For large-scale fading effects, we use the three-slope model as in \cite{CellFree_vs_SmallCell_hien2017}.
We train the GNN model on the data with different system settings, i.e., $N=[6,10,20]$ and $K = [20,30,40]$. The performance of the proposed distributed GNN-based framework is then evaluated through the testing data, which contains samples of different network settings from the training set. The trained GNN model includes $3$ MPGNN layers, and we set the hidden units of $\phi_1$ and $\phi_2$ as $\{12,16,32,64\}$ and $\{76,32,12,1\}$.  At the last layer, after using the ReLU activation function, we use the additional activation in \eqref{Activation fn: Power constraint} to render the final power allocation to satisfy the limited power.
\begin{figure}[t]
    \includegraphics[trim=3.4cm 8.5cm 1.5cm 8.5cm, clip=true, width=3.8in]{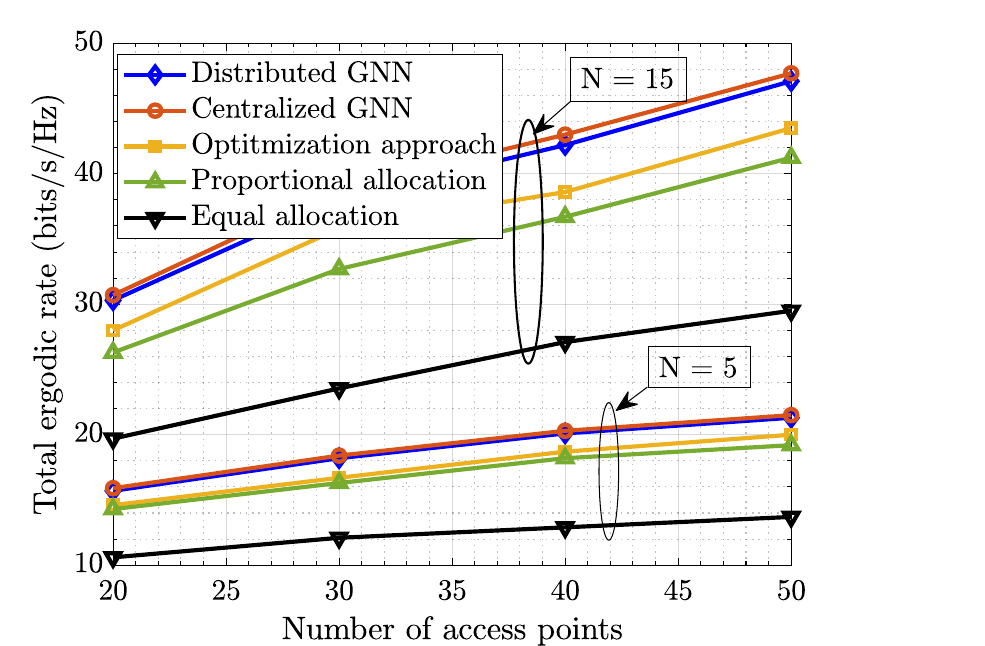}
    \caption{Total ergodic rate versus the number of APs when the system serves $N=5$ or $N=15$ UEs with $4$ antennas per AP.}
    \label{fig: Sumrate_vs_NoAPs}
\end{figure}

\begin{table}[t] 
    \scriptsize
	\caption{Total ergodic rate on seen network settings where the number of APs and users in the training and testing phases are the same.}
	\centering
	\begin{tabular}{|c|c|c|c|c|c|c|c|} 
		\toprule[1pt]\midrule[0.3pt] 
		M&N&K & $^{\text{Distributed}}_{\text{GNN}}$ & $^{\text{Centralized}}_{\text{GNN}}$ & $^{\text{Optimization}}_{\text{approach}}$ & $^{\text{Proportional}}_{\text{Allocation}}$ & $^{\text{Equal}}_{\text{Allocation}}$\\ 
		\midrule 
		4 & 6 & 30 & $\underset{\mbox{(109.2\%)}}{\mbox{21.5}}$ & $\underset{\mbox{(110.2\%)}}{\mbox{21.7}}$ & $\underset{\mbox{(100\%)}}{\mbox{19.7}}$ & $\underset{\mbox{(96.9\%)}}{\mbox{19.1}}$ & $\underset{\mbox{(69.1\%)}}{\mbox{13.6}}$ \\ 
		\midrule 
		4 & 10 & 40 & $\underset{\mbox{(109.3\%)}}{\mbox{33.1}}$ & $\underset{\mbox{(110.6\%)}}{\mbox{23.5}}$ & $\underset{\mbox{(100\%)}}{\mbox{30.3}}$ & $\underset{\mbox{(93.7\%)}}{\mbox{28.4}}$ & $\underset{\mbox{(69.6\%)}}{\mbox{21.1}}$ \\ 
		\midrule 
		10 & 6 & 30 & $\underset{\mbox{(109.7\%)}}{\mbox{27.1}}$ & $\underset{\mbox{(110.8\%)}}{\mbox{27.4}}$ & $\underset{\mbox{(100\%)}}{\mbox{24.7}}$ & $\underset{\mbox{(95.5\%)}}{\mbox{23.6}}$ & $\underset{\mbox{(70.4\%)}}{\mbox{17.4}}$  \\ 
		\midrule 
		10 & 10 & 40 & $\underset{\mbox{(108.6\%)}}{\mbox{42.9}}$ & $\underset{\mbox{(109.8\%)}}{\mbox{43.4}}$ & $\underset{\mbox{(100\%)}}{\mbox{39.5}}$ & $\underset{\mbox{(93.2\%)}}{\mbox{36.8}}$ & $\underset{\mbox{(69.6\%)}}{\mbox{27.5}}$ \\ 
		\bottomrule[1pt] 
	\end{tabular}
	\label{Table: Evaluate Seen data}
\end{table}%
\begin{table}[t] 
    \scriptsize
	\caption{Total ergodic rate on unseen network settings where the number of APs and users in the training and testing phases are different.}
	\centering
	\begin{tabular}{|c|c|c|c|c|c|c|c|} 
		\toprule[1pt]\midrule[0.3pt] 
		M&N&K & $^{\text{Distributed}}_{\text{GNN}}$ & $^{\text{Centralized}}_{\text{GNN}}$ & $^{\text{Optimization}}_{\text{approach}}$ & $^{\text{Proportional}}_{\text{Allocation}}$ & $^{\text{Equal}}_{\text{Allocation}}$\\ 
		\midrule 
		4 & 5 & 20 & $\underset{\mbox{(106.8\%)}}{\mbox{15.6}}$ & $\underset{\mbox{(107.5\%)}}{\mbox{15.7}}$ & $\underset{\mbox{(100\%)}}{\mbox{14.6}}$ & $\underset{\mbox{(95.2\%)}}{\mbox{13.9}}$ & $\underset{\mbox{(72.6\%)}}{\mbox{10.6}}$ \\ 
		\midrule 
		4 & 15 & 40 & $\underset{\mbox{(104.8\%)}}{\mbox{40.7}}$ & $\underset{\mbox{(107.3\%)}}{\mbox{41.5}}$ & $\underset{\mbox{(100\%)}}{\mbox{38.8}}$ & $\underset{\mbox{(93.3\%)}}{\mbox{36.2}}$ & $\underset{\mbox{(69.8\%)}}{\mbox{27.1}}$ \\ 
		\midrule 
		4 & 15 & 50 & $\underset{\mbox{(104.3\%)}}{\mbox{45.4}}$ & $\underset{\mbox{(105.5\%)}}{\mbox{45.9}}$ & $\underset{\mbox{(100\%)}}{\mbox{43.5}}$ & $\underset{\mbox{(92.6\%)}}{\mbox{40.3}}$ & $\underset{\mbox{(67.8\%)}}{\mbox{29.5}}$ \\ 
		\midrule 
		10 & 5 & 20 & $\underset{\mbox{(106.3\%)}}{\mbox{21.7}}$ & $\underset{\mbox{(107.4\%)}}{\mbox{21.9}}$ & $\underset{\mbox{(100\%)}}{\mbox{20.4}}$ & $\underset{\mbox{(94.1\%)}}{\mbox{19.2}}$ & $\underset{\mbox{(70.6\%)}}{\mbox{14.4}}$  \\ 
		\midrule 
		10 & 15 & 40 & $\underset{\mbox{(104.9\%)}}{\mbox{52.9}}$ & $\underset{\mbox{(105.5\%)}}{\mbox{53.2}}$ & $\underset{\mbox{(100\%)}}{\mbox{50.4}}$ & $\underset{\mbox{(92.4\%)}}{\mbox{46.6}}$ & $\underset{\mbox{(69.8\%)}}{\mbox{35.2}}$ \\ 
		\midrule 
		10 & 15 & 50 & $\underset{\mbox{(104.5\%)}}{\mbox{55.9}}$ & $\underset{\mbox{(106.0\%)}}{\mbox{56.7}}$ & $\underset{\mbox{(100\%)}}{\mbox{53.5}}$ & $\underset{\mbox{(92.0\%)}}{\mbox{49.2}}$ & $\underset{\mbox{(70.6\%)}}{\mbox{37.8}}$ \\ 
		\bottomrule[1pt] 
	\end{tabular}
	\label{Table: Evaluate Unseen data}
\end{table}%
\begin{table}[t] 
	\caption{Running time (second) of the considered benchmarks in the testing phase.}
	\centering
    \resizebox{\columnwidth}{!}{
	\begin{tabular}{|c|c|c|c|c|c|c|c|} 
		\toprule[1pt]\midrule[0.3pt] 
		N & K & $^{\text{Distributed}}_{\text{GNN}}$ & $^{\text{Centralized}}_{\text{GNN}}$ & $^{\text{Optimization}}_{\text{approach}}$ & $^{\text{Proportional}}_{\text{Alloc.}}$ & $^{\text{Equal}}_{\text{Alloc.}}$\\ 
		\midrule 
		6 & 20 & $0.0021$ & $0.0066$ & $536$ & $9.59 \times 10^{-6}$ & $5.88 \times 10^{-6}$ \\ 
		\midrule 
		10 & 20 & $0.0025$ & $0.0072$ & $1860$ & $9.63 \times 10^{-6}$ & $5.94 \times 10^{-6}$ \\ 
		\midrule 
		10 & 30 & $0.0025$ & $0.0079$ & $2950$ & $1.02 \times 10^{-5}$ & $6.10 \times 10^{-6}$ \\
		\midrule 
		20 & 30 & $0.0036$ & $0.0105$ & $-$ & $1.12 \times 10^{-5}$ & $6.86 \times 10^{-6}$ \\ 
		\midrule 
		20 & 50 & $0.0036$ & $0.0129$ & $-$ & $1.36 \times 10^{-5}$ & $7.95 \times 10^{-6}$ \\
		\bottomrule[1pt] 
	\end{tabular}
    }
	\label{Table: Computation Time}
\end{table}%

Fig.\ref{fig: Sumrate_vs_NoAPs} illustrates the performance of the proposed distributed GNN compared to given benchmarks. The total ergodic rate increases as more APs are added to the network. Despite the distributed GNN operating based on the local channel information, it yields results that are close to those of the centralized GNN in all the network settings. As the system serves $N=15$ UEs, the total ergodic rate produced by the distributed GNN-based framework increases from $29.1$~[bps/Hz] to $45.4$~[bps/Hz] if the number of APs increases from $K=20$ to $K=50$. Meanwhile, the centralized GNN improves the total ergodic rate from $29.5$~[bps/Hz] to $45.9$~[bps/Hz]. Both the centralized and distributed GNN frameworks outperform the optimization-based approach. The total ergodic rate produced by the optimization-based approach is consistently lower than the proposed distributed GNN, around $5-7$\%. The proportional allocation method also results in a lower ergodic rate compared to the distributed GNN by approximately $15-20$\%, although it performs better than the equal allocation method. The equal allocation method, while simple, yields the lowest ergodic rate among all methods. 

Tables \ref{Table: Evaluate Seen data} and \ref{Table: Evaluate Unseen data} show the adaptability of the proposed distributed GNN on different system settings, including scenarios with multiple antennas per AP. We observe that at both unseen and seen networks, the centralized GNN provides the total ergodic rate better than the proposed distributed GNN around $1$\% only for all the considered settings. The performance gap of both the learning-based approaches is reduced in comparison with the optimization approach on the unseen network settings, $M=[4,10]$, $K = [20, 40, 50]$ and $N=[5,15]$, from $9-10$\% to approximately $4-7$\%. The justification is obvious from the lack of information on different network sizes. To surmount this issue, we should train the GNN model utilizing diverse network sizes. Meanwhile, the effectiveness of the proportional allocation method decreases as network size increases, likely due to the need for more adaptive power distribution in larger networks.

Table \ref{Table: Computation Time} shows the running time of the three benchmarks. The runtime evaluation basically encompasses the graph construction and GNN response. In practice, the centralized GNN involves two additional steps, including uploading CSI to the CPU and broadcasting optimized power coefficients to APs, resulting in longer runtimes. The optimization-based approach exhibits the longest computation time due to an iterative manner. Thanks to basic arithmetic operators, the GNN-based frameworks process CSI quickly for prompt feedback. The distributed GNN framework exhibits a considerable advantage in running time compared to the centralized GNN as the number of APs increases. This is primarily because APs in the distributed scheme operate independently and in parallel, maintaining a consistent running time regardless of the number of APs. On the other hand, the centralized GNN requires aggregating data from all APs at the central CPU, leading to increased running time as the number of APs grows. The number of antennas does not affect the running time since it is included in the loss function only. The proportional and equal allocation methods have almost negligible running times due to their minimal computational requirements.
\section{Conclusion}\label{Conclusion}
In this paper, we have proposed a distributed GNN-based framework to allocate the transmit powers that maximize the sum ergodic rate in cell-free massive MIMO systems. Unlike the optimization-based and centralized GNN-based approaches, which are performed at the CPU and require CSI from all the APs, the proposed distributed GNN framework acquires local channel statistics to predict the power coefficients. The GNN model was trained at the CPU with local CSI from one AP, supplemented by partial information from other APs to capture the system's overall characteristics and share the computational workload among APs. The trained GNN model is then deployed locally at each AP in the testing phase. Numerical results demonstrated that the distributed GNN-based framework performs close to the centralized one and outperforms optimization-based approaches.

\bibliographystyle{IEEEtran}
\bibliography{Bib1}
\end{document}